%% file: symposium_paper.tex
\documentclass{iau} 
\usepackage{graphicx}
\usepackage{natbib}
\usepackage{amsmath}

\include{journal}

\newcommand{\msun}{\ensuremath{M_\odot}}
\newcommand{\lsun}{\ensuremath{L_\odot}}

\newcommand{\mdot}{\ensuremath{\dot{M}}}
\newcommand{\mej}{\ensuremath{M_\text{ej}}}

\newcommand{\teff}{\ensuremath{T_\text{eff}}}
\newcommand{\myr}{\ensuremath{\msun\,\text{yr}^{-1}}}
\newcommand{\mwind}{\ensuremath{M_\text{wind}}}

\title[Mass loss from binary stars approaching merger] %% give here short title %%
{Mass loss from binary stars approaching merger}

\author[O. Pejcha]   %% give here short author list %%
{Ondřej Pejcha}

\affiliation{Institute of Theoretical Physics, Faculty of Mathematics and Physics, Charles University, V~Holešovičkách 2, 180 00 Praha 8, Czech Republic \\ email: {\tt pejcha@utf.mff.cuni.cz}}

\pubyear{2022}
\volume{xxx}  %% insert here IAU Symposium No.
\jname{The Origin of Outflows in Evolved Stars}
\editors{L. Decin, A. Zijlstra, \& C. Gielen, eds.}
\begin{document}

\maketitle

\begin{abstract}
Some binary stars experience common envelope evolution, which is accompanied by drastic loss of angular momentum, mass, and orbital energy and which leaves behind close binaries often involving at least one white dwarf, neutron star, or black hole. The best studied phase of common envelope is the dynamical inspiral lasting few original orbital periods. We show theoretical interpretation of observations of V1309 Sco and AT2018bwo revealing that binaries undergo substantial prolonged mass loss before the dynamical event amounting up to few solar masses. This mass loss is concentrated in the orbital plane in the form of an outflow or a circumbinary disk. Collision between this slower mass loss and the subsequent faster dynamical ejection powers a bright red transient. The resulting radiative shock helps to shape the explosion remnant and provides a site of dust and molecule formation.
\end{abstract}

\firstsection % if your document starts with a section,
\section{Introduction}

Many binary stars undergo at least one episode of common envelope (CE) evolution. This short evolutionary phase causes ejection of a considerable fraction of total binary mass, significantly reduces the orbital separation of surviving bodies, or leads to a merger of the two binary components \citep[e.g.][]{paczynski76,iben93,sana12,ivanova13}. CE evolution is important for the formation of many objects of astrophysical importance, including gravitational wave sources \citep[e.g.][]{dominik12}.

The binary star typically starts CE by developing a phase of unstable mass transfer. As the mass transfer rates gradually increase, the accreting star cannot accept this inflow of material and a fraction of the mass leaving the donor likely escapes the binary system altogether. Most of this material leaves the binary in the vicinity of Lagrange points L2 or L3. As the mass transfer instability runs away, the fraction of mass leaving the binary increases. Similar outcome likely occurs when the two stars begin their spiral-in due to the tidal Darwin instability. Eventually, the evolution of the two stars becomes fully dynamical, which can be viewed as an instantaneous ejection of material. The surviving binary or single merged star then relaxes to hydrodynamical and thermal equilibrium. 

Traditionally, CE has been studied by comparing pre- and post-CE populations of binary and single stars. New discoveries and increasing volume data from time-domain transient surveys have opened new ways how to study CE evolution. In particular, a class of transients named Luminous red novae (LRNe) is now associated with CE events \citep{ivanova13_sci}. In this contribution, we discuss astrophysical interpretations of time-series observations before and during the merger. We study the possible outcomes in low- and high-mass stars by interpreting observations of V1309~Sco and AT2018bwo, respectively.

\section{Gradual mass loss preceding dynamical phase}

Recent binary evolution models suggest that the runaway binary mass transfer can last many hundreds or thousands of orbits and that mass-loss  rates can exceed $\mdot \gtrsim 10^{-2}\,\myr$ \citep{blagorodnova21}. Much of the gas from the donor leaves the binary altogether. By studying trajectories of ballistic test particles leaving the L2 point, \citet{shu79} showed that the tidally-torqued gas either leaves to infinity or forms a circumbinary disk. \citet{hubova19} found a wider varied range of outcomes when they considered particles with initial kicks or positional offsets from L2.  \citet{pejcha16a,pejcha16b} studied the radiative hydrodynamics of the same problem. They found that as the spiral stream expands, the spiral windings collide with themselves forming radial internal shocks. The velocity difference in the shocks $\Delta v$ is closely related to the binary orbital velocity, $\Delta v \propto \sqrt{GM/a}$, where $M$ is the binary mass. The resulting shock is radiative and powers emission with the luminosity of the order of $L \propto \mdot (\Delta v)^2$. For high $\mdot$, the outflow is optically-thick and the shock power is adiabatically degraded before it can radiate. 

Ignoring viewing-angle effects, the L2 outflow is an additional source of light added on top of the central binary star. Depending on the binary and mass-loss properties, we can expect two possible outcomes. When the L2 outflow dominates, we should observe a gradual increase of $L$ before the main outburst, although the emission from the L2 outflow might come out at mostly in the infrared. If the central binary dominates, we might observe constant pre-outburst flux or even dimming of the central binary due to dust obscuration by the outflow. 

What is the dividing line between the two regimes? We can express $\mdot \sim M/P$, where $P$ is the orbital period, and combine it with Kepler's laws to estimate L2 luminosity as $L \sim \mdot (\Delta v)^2 \sim M^2/(Pa) \sim M^{2.5}/a^{2.5}$. This approximation is very crude, because $\mdot$ is likely much smaller than $M/P$ for most of the pre-CE evolution and because the actual value of $\mdot$ is set by the structure of the mass-losing star and binary properties. For a Roche-lobe filling primary star on the main sequence, $a \sim M^{0.8}$, which gives $L \sim M^{0.5}$. This implies that L2 luminosity grows only very slowly with the binary mass and is weaker for more evolved primaries. Luminosity of the stars on the main sequence scales as $L \propto M^{3.5}$ and the luminosity of more evolved stars of the same mass is even higher. This means that the effect of L2 mass loss will be harder to detect in high-mass binaries. 

\underline{\it V1309 Sco} was classified as a LRN by \citet{mason10}. \citet{tylenda11} analyzed dense photometric dataset from the \textit{OGLE} survey  covering approximately 7~years before the explosion and found that V1309~Sco was initially a contact binary with $P\approx 1.4$\,days. Orbital period experienced rapid decrease on the approach to the merger, which was accompanied by change of the orbital light curve from double-hump to single-hump profile. About a year before the peak brightness, the orbital variability disappeared, the observed flux shortly decreased and then gradually increased to the main peak.

\citet{pejcha17} used semi-analytic models combined with smoothed particle hydrodynamic simulations with flux-limited diffusion treatment of radiation in the vertical direction of the equatorially-concentrated outflow to explain the observed pre-explosion behavior of V1309~Sco. They explained the change of orbital light curve profile by setting the inclination angle of the binary to about $80^\circ$ and viewing it through an L2 spiral outflow with gradually increasing $\mdot$. Mass leaving the binary also carries angular momentum, which leads to the decrease of $P$. \citet{pejcha17} showed that $\mdot$ inferred from changing light curve shape and $\dot{P}$ inferred from orbital period variations broadly agree with each other. 

For $\mdot \gtrsim 10^{-3}\,\myr$, the binary is obscured by the L2 outflow and the orbital variability was not visible anymore. After this, internal shocks in the L2 outflow provide enough luminosity to increase the observed $L$. Since the outflow is optically-thick, interplay of diffusion and adiabatic expansion control the amount of released radiation; \citet{pejcha17} found reasonably good match to the observed light curve for a prescribed evolution of $\mdot$ with numerical simulations. \citet{pejcha17} also found indications that the properties of the outflow change approximately 50\,days before the merger, which can be caused either by growing temperature of the binary due to mass-loss stripping or changes to the morphology of the mass loss flow \citep{macleod18}.

\begin{figure}[t]
% \vspace*{-2.0 cm}
\begin{center}
 \includegraphics[width=\textwidth]{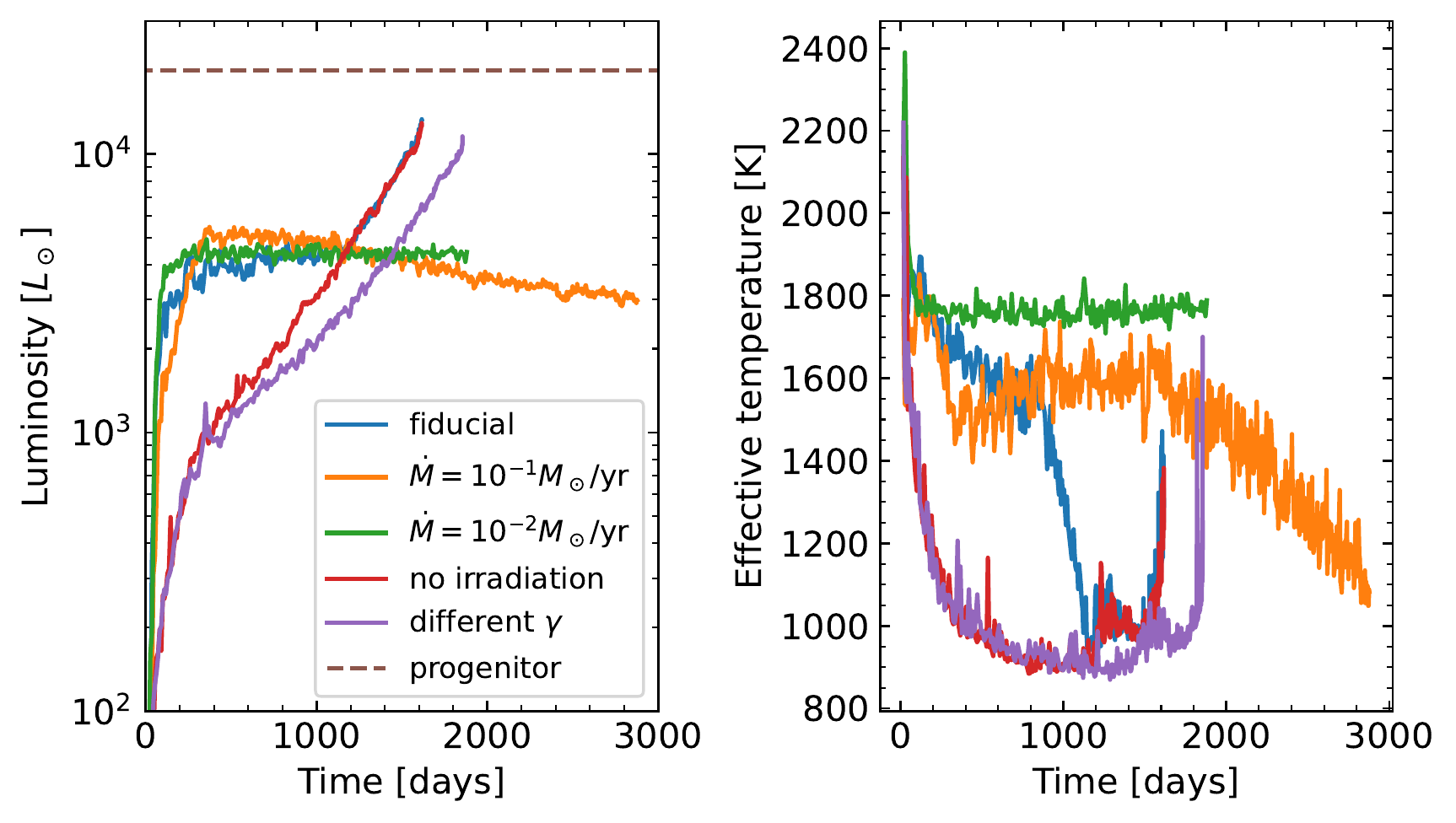} 
% \vspace*{-1.0 cm}
 \caption{Radiative properties of mass ejected from L2 in a binary modeled after AT2018bwo \citep{blagorodnova21}. We show the evolution of luminosity (left panel) and mean effective temperature (right panel). The simulation was performed with the smoothed particle hydrodynamics code with vertical radiative diffusion presented in \citet{pejcha16a,pejcha16b,pejcha17}, but for a binary with $M_1=2.6\,\msun$, $M_2=13\,\msun$, $a=1$\,AU, and the binary effective temperature $\teff = 6000$\,K, which gives binary luminosity assumed in the code $L=4\pi a^2 \sigma \teff^4 \approx 5 \times 10^4\,\lsun$. The mass loss rate of the fiducial model (blue lines) was initially set to $\mdot = 3\times 10^{-2}\,\myr$, which increased as a power law with an index $\gamma=3$ and with a singularity set to $t = 2000$\,days. We also show modifications of the fiducial model by setting $\mdot$ constant (orange and green lines), no irradiation by the central binary (red lines), and $\gamma=2$ (purple lines). For comparison, dashed vertical line in the left panel shows the progenitor luminosity of AT2018bwo, $L\approx 2\times 10^4\,\lsun$ observed about 14 years before the outburst by \citet{blagorodnova21}. }
   \label{fig:1}
\end{center}
\end{figure}

\underline{\it AT2018bwo} was thoroughly analyzed by \citet{blagorodnova21} using a combination of pre-explosion photometry, spectroscopy, binary evolution models, and modeling of the transient. They found that the progenitor position in the Hertzprung-Russel diagram matches $M\approx 15\,\msun$ binary with $a\approx 1$\,AU undergoing thermal-timescale mass transfer with $\mdot \approx 10^{-2}\,\myr$. 

In Figure~\ref{fig:1}, we show the results of modeling AT2018bwo using the similar assumptions as was done for V1309~Sco; more thorough discussion is in Section~4.4 of \citet{blagorodnova21}. We see that even under optimistic assumptions the L2 luminosity does not reach the luminosity of the progenitor observed approximately 14 years before the peak of the outburst. This is different from V1309~Sco, where the L2 outflow was the dominant source of luminosity for a year before the merger, but it is also expected based on our analytic estimates.

Figure~\ref{fig:1} also shows estimates of effective temperature $\teff$ of the L2 radiation. We see that the expected value is between 1000 and 2000\,K, which implies that most of the luminosity will be seen in the near infrared. However, it is not clear whether the outflow is sufficiently cool for dust condensation.

\section{Collision of dynamical ejecta with pre-explosion mass loss}

\begin{figure}[t]
% \vspace*{-2.0 cm}
\begin{center}
 \includegraphics[width=\textwidth]{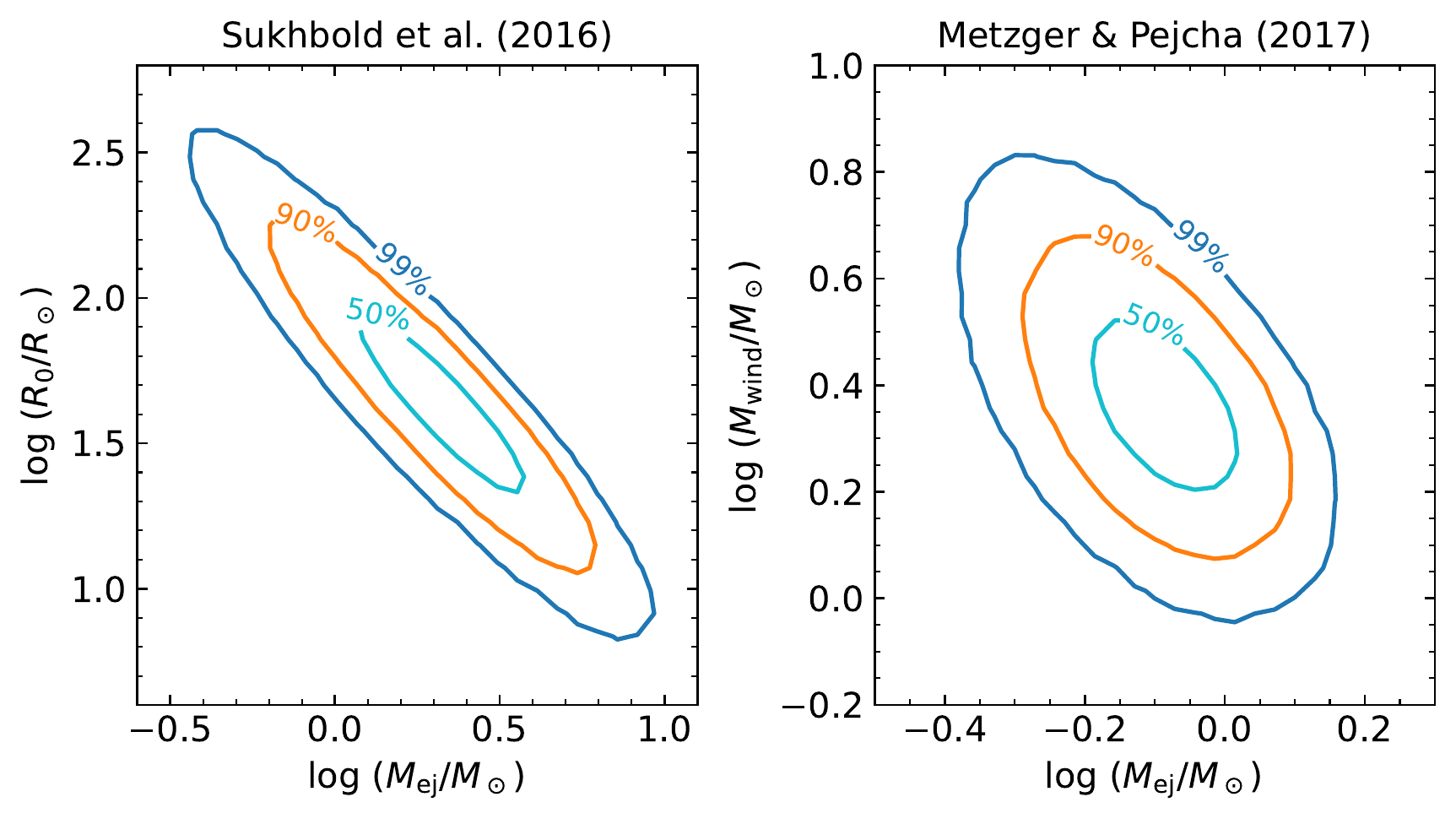} 
% \vspace*{-1.0 cm}
 \caption{Ejecta properties of AT2018bwo estimated from its plateau luminosity, duration, and expansion velocity. The left panel shows the confidence ellipsoids for the initial radius $R_0$ and ejecta mass $\mej$ in the model of the scaled-down Type II-P supernova \citep{sukhbold16}. The right panel shows the confidence ellipsoids for the mass of pre-explosion outflow $\mwind$ and mass of the ejecta $\mej$ in the shock collision model \citep{metzger17}.}
   \label{fig:2}
\end{center}
\end{figure}

Once the timescale of the acceleration of the binary inspiral becomes shorter than the expansion timescale of the outflow near the binary, it becomes more convenient to think about the subsequent mass loss as a nearly instantaneous mass ejection. The ejecta likely moves faster than the previous L2 outflow, because it was launched from a binary on a much tighter orbit, and it is also likely less concentrated within the equatorial plane, because of more shock heating. As the more spherical faster ejecta expands, it will radiate part of its thermal energy. Envelopes of most stars are hydrogen-rich and the resulting transient should resemble a scaled-down version of Type II-P supernova, as was first pointed out by \citet{ivanova13_sci}. Multiple peaks in the light curves of LRNe are explained as individual mass ejections. 

It is perhaps inevitable that the faster more spherical ejecta collides with the pre-existing equatorial outflow forming a radiative shock. The hydrodynamics of such an interaction are relatively well understood \citep[e.g.][]{suzuki19,kurfurst19,kurfurst20,mcdowell18}, but the implications for transients are less explored. \citet{metzger17} constructed a semi-analytic model of an equatorial radiative shock coupled to an expanding envelope. They argued that the first peak in the light curves of LRNe comes from cooling emission from the freely-expanding polar ejecta \citep{macleod17}, while the second peak is caused by diffusion of light from the dense equatorial radiative shock. The reported scaling relations can explain luminosities and timescales of the observed events and suggest that some of the  long infrared transients recently identified  by \citet{kasliwal17} are CE events from evolved binaries on wide orbits.

\underline{\it V1309~Sco} shows a single peak, which can be explained by ejecting few hundredths of $\msun$ of recombining hydrogen \citep{ivanova13_sci,nandez14}. In the shock-powered model, the second peak can be hidden behind the dust formed near the radiative shock in the equatorial plane, because we are viewing the system near the original orbital plane. Alternatively, the shock might not be energetic enough to keep the hydrogen ionized for sufficiently long \citep{metzger17}.

\underline{\it AT2018bwo} also showed a single peak, but the data are substantially scarcer than in V1309~Sco. Bolometric light curve of \citet{blagorodnova21} shows a possible brightening toward the end of the plateau, potentially resembling a second peak. The transient properties were analyzed by \citet{blagorodnova21} using analytic scaling relations in the Type II-P supernova and shock-powered models.  They found that different Type II-P supernova scaling relations give very different inferences of the ejection radius $R_0$ and ejecta mass $\mej$, because their application to LRNe is an extrapolation from the domain where they have been validated by radiation hydrodynamics simulations. The analytic scalings of the shock powered model of \citet{metzger17} give reasonable values for the masses of the pre-existing equatorial outflow ($\mwind$) and of the faster ejecta ($\mej$). The inference  shows that $\mwind > \mej$, which is in agreement with binary evolution models of the same event of \citet{blagorodnova21}. In Figure~\ref{fig:2}, we show the confidence ellipsoids of the inferred physical parameters for the two models of the transient.

\section{Future outlook}

Observations of LRNe can provide new insight into the open questions in the CE evolution. We have argued that the most often studied dynamical phase of CE evolution is preceded by a long gradual loss of mass from the binary, which can be observed as a slow rise of brightness. When the event becomes dynamical, the faster younger and more spherical ejecta should collide with the older equatorially-concentrated mass distribution. The resulting radiative shock can explain the observed luminosities and timescales as well as double peaks seen in some events.

But there remains much to be done. A predictive theory of mass-loss rate evolution before the dynamical phase remains to be found. Standard spherically-symmetric stellar evolution codes can be evolved far enough to give very high mass-loss rates, but they currently cannot reach close enough the main peak. Dust formation in the gradual equatorial outflow and its observational signatures need to be properly characterized. Modeling of the transients would benefit from calibrating the analytic scaling relation of Type II-P supernovae with radiation hydrodynamic simulations appropriate for LRNe. The shock powered model needs to be more developed to be directly comparable to observations. This is difficult, because the problem geometry deviates from spherical symmetry and it is necessary to include realistic equation of state as well as appropriate opacities and take into account dust formation.

\section*{Acknowledgements}

This research was supported by Horizon 2020 ERC Starting Grant ‘Cat-In-hAT’ (grant agreement no. 803158).

\end{document}

%% file: journal.tex
\def\apj{ApJ}%
\def\aap{A\&A}%
\def\aapr{A\&A~Rev.}%
\def\mnras{MNRAS}%
\def\pasp{PASP}%